\documentclass[twocolumn, longbibliography, prl, amsmath, amssymb,superscriptaddress]{revtex4-2}
\usepackage{amsfonts}
\usepackage{amsmath}
\usepackage{graphicx}
\usepackage[colorlinks,linkcolor=blue,citecolor=blue,urlcolor=blue]{hyperref}
\usepackage{float}
\usepackage[normalem]{ulem}
\usepackage[T1]{fontenc}
\usepackage{lipsum}
\usepackage[ampersand]{easylist}
\usepackage{comment}
\usepackage{amsmath,amssymb,graphicx}
\usepackage{soul,xcolor}
\usepackage{textcomp}
\usepackage{siunitx}
\usepackage{xspace}
\usepackage{orcidlink}

\makeatletter
\renewcommand{\fnum@figure}{\textbf{Fig.~\thefigure~|}}
\renewcommand{\@caption@fignum@sep}{\enspace}
\makeatother

\begin{document}

\title{Limit Cycles in a Photonic Dimer with Tuneable Non-Hermitian Interactions}

\author{K. J. H. Peters\,\orcidlink{0000-0002-0626-6446}}
\email{kpeters4@uni-bonn.de}
\affiliation{Institut f\"ur Angewandte Physik, Universit\"at Bonn, Wegelerstrasse 8, 53115 Bonn, Germany}

\author{P. Schnorrenberg,\orcidlink{0009-0002-5183-6361}}
\affiliation{Kirchhoff-Institut für Physik, Im Neuenheimer Feld 225a, 69120 Heidelberg, Germany}

\author{D. Ehrmanntraut\,\orcidlink{0000-0003-1993-2775}}
\affiliation{Kirchhoff-Institut für Physik, Im Neuenheimer Feld 225a, 69120 Heidelberg, Germany}

\author{N. Longen\,\orcidlink{0009-0007-9590-0181}}
\affiliation{Kirchhoff-Institut für Physik, Im Neuenheimer Feld 225a, 69120 Heidelberg, Germany}

\author{J. Schmitt\,\orcidlink{0000-0002-0002-3777}} 
\email{julian.schmitt@kip.uni-heidelberg.de}
\affiliation{Institut f\"ur Angewandte Physik, Universit\"at Bonn, Wegelerstrasse 8, 53115 Bonn, Germany}
\affiliation{Kirchhoff-Institut für Physik, Im Neuenheimer Feld 225a, 69120 Heidelberg, Germany}

\begin{abstract}
Interactions govern the emergence of collective behaviour in classical and quantum many-body systems. While conservative interactions are well known to generate nonlinear phenomena ranging from self-trapping to pattern formation, it remains largely unexplored whether purely dissipative -- i.e., non-Hermitian -- interactions can give rise to similarly rich dynamics and nontrivial system states. Here, we experimentally realise tuneable non-Hermitian interactions in two coupled condensates of light confined within a dye-filled double-well microcavity. Local coupling to molecular reservoirs generates the effective dissipative photon interactions. We show that the interplay between coherent tunnelling and interactions stabilises limit-cycle oscillations, a hallmark of nonlinear dynamics traditionally associated with Hermitian nonlinearities. By tuning the reservoir coupling, we map out the dynamical phase diagram comprising stable fixed points and limit cycles, thereby demonstrating direct control over the interaction strength. Our experimentally validated model reveals both supercritical and subcritical Hopf bifurcations, giving rise to hysteresis, bistability and excitability. These results validate dissipative interactions as a mechanism for organising collective nonlinear dynamics in driven-dissipative systems and pave the way towards exploring nonequilibrium many-body physics through controlled dissipation.
\end{abstract}
\date{\today}
\maketitle


\section*{Introduction}
Collective many-body states and nonequilibrium attractors emerge in a wide range of physical systems, from strongly correlated electron systems to ultracold gases~\cite{Lee:2006,Bloch:2012}. In closed quantum systems, emergent phenomena and nonlinear dynamical phases, such as superfluidity~\cite{Leggett:1999} or macroscopic self-trapping~\cite{Albiez:2005,Abbarchi:2013}, arise from the interplay of kinetic motion and interparticle interactions, which give rise to an effective density-dependent Hermitian (\textit{i.e.}, real-valued) nonlinearity. Interactions thus act as an effective mean-field potential $ g|\psi(\mathbf{r})|^2$, whose spatial gradients drive the flow of the particle density $|\psi(\mathbf{r})|^2$ with strength $g$.

In open quantum systems, in contrast, coupling to the environment introduces dissipation and gain, which leads to non-Hermitian effects; notable examples that have been extensively studied in recent years include exceptional points, topological states, or non-reciprocal transport~\cite{ElGanainy:2018,Miri:2019,Ashida:2020}. Beyond linear non-Hermitian effects, density-dependent nonlinear couplings to the environment can also generate effective dissipative (\textit{i.e.}, imaginary-valued) interactions, with gain saturation in lasers or two-photon absorption constituting well known examples~\cite{Siegman:1986,Malaguti:2011}. In such non-conservative systems, particle transport is instead governed by gradients of an imaginary potential $i\mathcal{G}|\psi(\mathbf{r})|^2$ of strength $\mathcal{G}$, which generate local gain and loss rather than conservative forces. This formal analogy raises the fundamental question whether systems with local dissipative interactions can organise nonlinear attractor landscapes in the same way that conservative interactions organise collective states.

A direct way to address this question is through self-sustained oscillations, which arise when a stationary state loses stability and evolves towards a limit cycle~\cite{Strogatz:2024, Jenkins:2013}. As stable nonequilibrium attractors, limit cycles provide a direct signature of nonlinear interactions organising the dynamical phase space of an open system. They occur across a broad range of nonlinear classical and quantum systems, from electronic and mechanical oscillators to atomic gases, chemical reactions and biological dynamics~\cite{Jenkins:2013,Winfree:1967,Kongkhambut:2022,Zhabotinsky:1991}. In optics, limit cycles have been observed in lasers and nonlinear optical cavities~\cite{Orozco:1984,Metzger:2008,Chen:2012,Brunstein:2012,CarlonZambon:2020}. To date, however, these observations have relied on real-valued Hermitian nonlinearities, including Kerr interactions~\cite{CarlonZambon:2020,Marconi:2020,Kim:2020}, or on slow feedback mechanisms~\cite{Navarro-Urrios:2017,Abad-Arredondo:2024,Peters:2026}, and their stabilisation solely by dissipative interactions has so far not been observed.

To resolve this challenge experimentally requires independent control over coherent transport and dissipative interactions. Photon and exciton–polariton condensates in dye- or semiconductor-filled microcavities provide precisely such a platform~\cite{Klaers:2010,Bloch:2022,Schofield:2024,Pieczarka:2024}. Their kinetic energy can be engineered through tunnel-coupled cavities and photonic lattices~\cite{Jacqmin:2014,Dung:2017,Redmann:2024}, while coupling to molecular or excitonic reservoirs naturally generates effective non-Hermitian interactions through thermalisation, gain and loss, and reservoir dynamics~\cite{Gladilin:2020b,Abouelela:2025}. These ingredients have enabled observations of non-Hermitian phase transitions, nonlinear relaxation dynamics and universal scaling~\cite{Oeztuerk:2021,Sazhin:2024,Erglis:2025,Fontaine:2022,Widmann:2026}. Yet, dissipative interactions themselves have so far not been experimentally accessible as independently tuneable, local interaction parameters.

Here we report the observation of limit-cycle dynamics arising purely from dissipative photon interactions. The experiment is realised in two photon condensates coherently coupled by a tunnel junction inside a dye-filled double-well microcavity, forming a photonic dimer in which local coupling to molecular reservoirs mediates on-site interactions. Gain saturation and reservoir dynamics give rise to a fast non-Hermitian interaction whose strength and sign are experimentally tuneable. In contrast to bosonic Josephson junctions, where coherent tunnelling and conservative interactions give rise to Josephson oscillations and self-trapping~\cite{Albiez:2005,Abbarchi:2013}, the dissipative interactions reported here stabilise nonlinear attractors, including limit cycles, bistability and hysteresis. Our results thereby demonstrate that non-Hermitian interactions can organise nonlinear attractor landscapes in many-body systems.

\section*{Results}
\subsection*{Model and Experimental Scheme}
\begin{figure}[t]
    \includegraphics[width=\columnwidth]{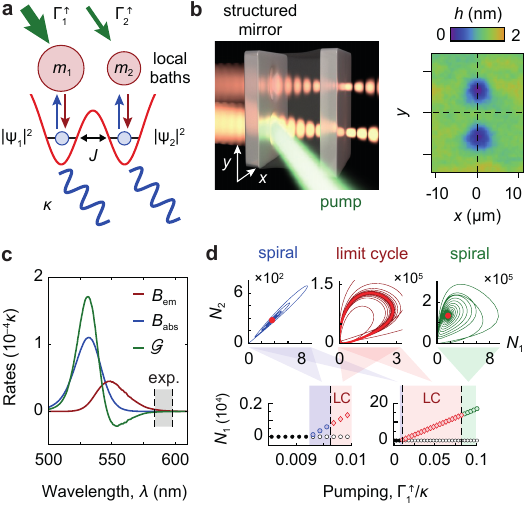}
    \caption{\textbf{Experimental scheme.} (a)~Driven-dissipative photonic dimer model consisting of two tunnel-coupled condensate states $\psi_{1,2}$ with effective interactions mediated by coupling to local molecule baths $m_{1,2}$. (b)~Dye-filled optical microcavity (left) and height profile of structured mirror (right). (c)~Measured wavelength-dependent absorption (blue), emission (red) spectra, and calculated effective interaction strength (green) for $\Gamma_1^\uparrow=2\pi\times\SI{200}{\mega\hertz}$. The studied experimental range is shaded. (d)~Calculated steady-state photon numbers versus pump rate $\Gamma_1^\uparrow$ (bottom panel). Black circles: stable fixed points below condensation; blue/green circles: stable fixed points above condensation; open black circles: single-mode unstable; red diamonds: parametrically unstable domain with limit cycles (LC). The stability domains are separated by Hopf bifurcations (dashed lines). Top panels show numerical simulations in the $\left(N_1,N_2\right)$-plane of left and right well population. Parameters: $\Gamma_2^\uparrow=0$, $\Gamma^\downarrow=2\pi\times\SI{250}{\mega\hertz}$, $M=4.4\times 10^6$, $J=2\pi\times\SI{3.6}{\giga\hertz}$, $\lambda_c=\SI{588}{\nano\meter}$.} \label{fig:1}
\end{figure}

Figure~\ref{fig:1}(a) outlines our experimental scheme. We consider two photon condensates confined in a double-well potential. The photons tunnel between the sites and couple locally to pumped molecular reservoirs by absorption and emission processes. Under balanced pumping, Bose-Einstein condensation occurs in the symmetric ground state~\cite{Kurtscheid:2019}. In the present work, we pump excitations into the system via the reservoirs in a highly site-selective way, while photon losses remain spatially uniform. This allows us to prepare nonequilibrium initial conditions and investigate the reservoir-induced effective photon interactions by monitoring emergent limit-cycle dynamics.

Experimentally, we utilise a dye-filled microcavity formed by highly-reflective mirrors, of which one has two concave surface features spaced by $\SI{10.9\pm0.1}{\micro\meter}$ imprinted by a structuring method~\cite{Kurtscheid:2020,Vretenar:2023}, as shown in Fig.~\ref{fig:1}(b). The surface height variation of $\SI{1.4\pm 0.1}{\nano\meter}$ realises a trapping potential of depth $U_\mathrm{D}/\hbar=2\pi\times \SI{504\pm36}{\giga\hertz}$ in frequency units, while the measured photon tunnelling rate resulting from evanescent overlap of the two Gaussian modes is $J=2\pi\times\SI{3.6\pm 0.1}{\giga\hertz}$ (see Methods). Due to the short cavity length $D_0\approx \SI{1.4}{\micro\meter}$, the free spectral range ($\SI{73}{\tera\hertz}$) exceeds the spectral bandwidth of the dye molecules ($\SI{6.2}{\tera\hertz}$)~\cite{Klaers:2010}, so that the longitudinal mode number of the photons remains fixed during the experiment. Effectively, this introduces a low-frequency cutoff at the wavelength $\lambda_\mathrm{c}$ and restricts the dynamics to the transverse plane where photons populate two hybridised states in the double well. As shown in Fig.~\ref{fig:1}(b), the molecules are optically pumped predominantly at one of the two sites using a laser beam at $\SI{532}{\nano\meter}$, which is close to the dye absorption maximum [see Fig.~\ref{fig:1}(c)]. 


We model the system by Langevin equations for the photon fields $\psi_j$
\begin{equation}
\label{eq:model}
 i \hbar \begin{pmatrix} \dot{\psi}_1 \\   \dot{\psi}_2 \end{pmatrix}
 = \hbar \begin{pmatrix}  \tilde\Delta_1 & J  \\  J & \tilde\Delta_2  \end{pmatrix} \begin{pmatrix} {\psi}_1 \\   {\psi}_2 \end{pmatrix}
 + i\hbar \begin{pmatrix}  \sqrt{D(m_1)}\;\xi_1    \\     \sqrt{D(m_2)}\;\xi_2 \end{pmatrix},
\end{equation}
coupled to rate equations $\dot{m}_j=\Gamma_j^\uparrow \left( 1-m_j \right) - \Gamma^\downarrow m_j + \left[B_\textrm{abs}(1-m_j) - B_\textrm{em} m_j\right] |\psi_j|^2$ for the excited molecule fractions at the two sites (see Methods). The index $j=1,2$ denotes the strongly- and weakly-pumped well, respectively. Here, $\Gamma_j^\uparrow$ is the site-dependent pump rate, $\Gamma^\downarrow$ non-radiative molecule decay, $B_\textrm{abs,em}(\lambda)$ the absorption and emission rates, $\kappa$ the cavity loss rate, $M$ the total number of molecules per site, and $J$ the coherent intercavity coupling [Fig.~\ref{fig:1}(a)]. In Eq.~\eqref{eq:model} we have introduced the effective detuning $\tilde\Delta_j=\Delta_j + i \left[-\kappa - B_\textrm{abs}M\left(1-m_j\right) + B_\textrm{em}M m_j\right]/2$, which depends on the dye-cavity detuning $\Delta_j=\omega_j-\omega_\mathrm{zpl}=2\pi c(\lambda_j^{-1}-\lambda_\mathrm{zpl}^{-1})$ between the condensate frequencies $\omega_j$ and the zero-phonon line $\omega_\mathrm{zpl}$ and the coupling to the local environment. Finally, we include noise of magnitude $D(m_j) =  \left[\kappa + B_\textrm{abs}M\left(1-m_j\right) + B_\textrm{em}M m_j\right]/2$ by complex Gaussian processes $\xi_j(t)$ with $\left\langle \xi_j(t)\right\rangle=0$ and $\left\langle \xi_j(t)\xi_k^*(t')\right\rangle=\delta_{jk}\delta(t-t')$ to account for spontaneous emission.

The photon-molecule coupling in Eq.~\eqref{eq:model} mediates an effective non-Hermitian interaction between the photons. To illustrate this explicitly, we adiabatically eliminate the molecular dynamics by setting $\dot m_j=0$, which is well justified for weak pumping as the photon-molecule dynamics are slow compared to the reservoir relaxation rates (see Methods). One obtains a nonlinearity that acts purely through gain and loss, $i\hbar \dot\psi_j = (\hbar\tilde\Delta'_j + i\hbar \mathcal{G}_j|\psi_j|^2) \psi_j - \hbar  J\psi_{3-j}$. Here, the first term represents a modified linear on-site energy and gain (see Methods), while the second term constitutes a density-dependent non-Hermitian interaction of strength
\begin{equation}\label{eq:interaction}
    \mathcal{G}_j=\frac{M\left(B_\textrm{abs}+B_\textrm{em}\right)}{2}\frac{B_\textrm{abs} \Gamma^\downarrow-B_\textrm{em}\Gamma^\uparrow_j}{(\Gamma^\downarrow+\Gamma^\uparrow_j)^2},
\end{equation}
which depends on the local pump rate and the wavelength-dependent absorption and emission rates. Thus, $\mathcal{G}_j$ can be experimentally tuned, \textit{e.g.}, by varying the cavity length $D_0$, which sets the cutoff wavelength $\lambda_\mathrm{c}$. Figure~\ref{fig:1}(c) shows the measured absorption and emission spectra~\cite{Schmitt:2024}, together with $\mathcal{G}_j$, which varies in magnitude and sign as $\lambda_\mathrm{c}$ is tuned.

Figure~\ref{fig:1}(d) shows calculated steady-states and photon number dynamics for the driven-dissipative  photonic dimer. As the pump rate $\Gamma^\uparrow_1$ increases, the steady-state population $N_1=|\psi_1|^2$ grows and the nonlinear system undergoes a sequence of stability changes (see Methods). First, upon reaching condensation, a finite population emerges to which the system relaxes with spiral dynamics (blue data). A second transition occurs at a Hopf bifurcation, where the stationary steady-state loses stability and limit cycles appear, visible as self-sustained oscillations (red data). At even larger $\Gamma^\uparrow_1$, they vanish at another Hopf bifurcation (green data). We have verified that the Hopf bifurcations predicted by the effective interaction picture closely match the exact photon--reservoir model (see Methods). Note that while the linear stability analysis shown in the bottom panels of Fig.~\ref{fig:1}(d) resembles that of parametric instabilities in coherently-driven polariton  or Kerr-nonlinear cavity systems~\cite{Sarchi:2008,CarlonZambon:2020}, the microscopic origin of the here reported oscillatory dynamics arises from dissipative, reservoir-mediated interactions and is therefore fundamentally different.


\begin{figure}[t]
   \includegraphics[width=\columnwidth]{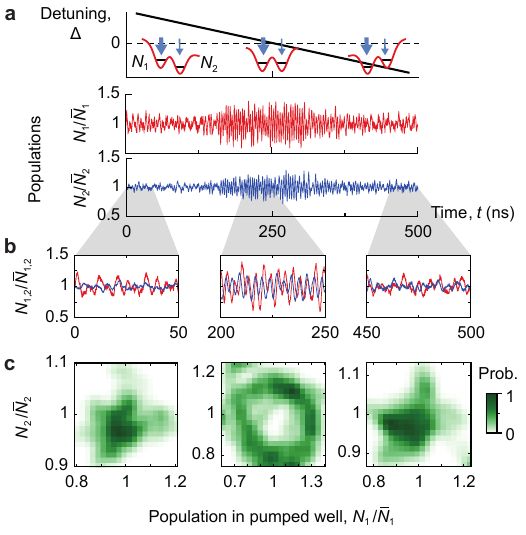}
 \caption{\textbf{Observation of limit cycles in double well.} (a)~Top: Sketch of the experimental detuning evolution $\Delta(t)$. Starting from an initially blue-detuned double well ($t<\SI{250}{\nano\second}$), the two sites are shifted on resonance $\Delta=0$, before entering the red-detuned regime ($t>\SI{250}{\nano\second}$). Middle and bottom: Single-shot measurement of the normalised photon numbers at constant pump power $P_1=\SI{3.42}{\micro\watt}$ and cutoff $\lambda_c=\SI{588}{\nano\meter}$. Red (blue) trace shows the emission from the pumped (unpumped) well $N_{1(2)}/\bar{N}_{1(2)}$. (b)~Zooms into the regions indicated in (a). While for $|\Delta|>0$, a steady-state emission without periodic features is observed, limit cycle dynamics are observed near $\Delta= 0$. (c)~The corresponding histograms in the photon population $\left(N_1,N_2\right)$-plane reveal the limit cycle. The histograms were obtained from 100 single-shot trajectories of $\SI{20}{\nano\second}$ each, corresponding to $\sim 700$ oscillation periods. }
 \label{fig:2}
\end{figure}

\begin{figure*}[t]
    \includegraphics[width=1.0\textwidth]{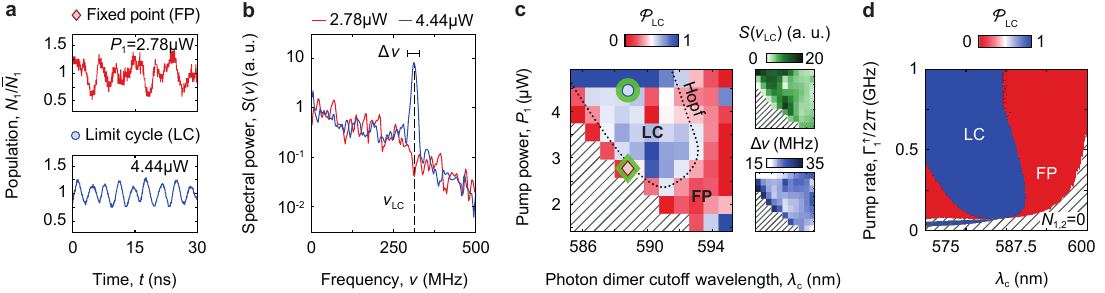}
    \caption{\textbf{Phase diagram of the nonlinear photonic dimer dynamics.} (a)~Time traces of the normalised population in the left well $N_1/\bar{N}_1$ for a stable fixed point (top) and limit cycle (bottom) for two pump powers at resonance $\Delta=0$. The noise levels differ due to different $\bar N_1$. (b)~Corresponding power spectral densities $S(\nu)$. For the limit cycle, a peak at frequency $\nu_\mathrm{LC}=\SI{313.1\pm0.2}{\mega\hertz}$ with width $\Delta\nu=\SI{13.6\pm0.5}{\mega\hertz}$ (FWHM) is visible. (c)~Limit cycle probability $\mathcal{P}_\mathrm{LC}$ (main panel), average power spectral density $S(\nu_\mathrm{LC})$ at the limit cycle frequency (top right, green), and $\Delta\nu$ (bottom right, blue) versus cutoff wavelength and pump power. Limit cycles occur in an island (blue region) well separated from the stable fixed point region (red) by a Hopf bifurcation (dashed line is a guide to the eye). We have verified that the phase boundaries are well preserved when varying the criteria for $\Delta \nu$ by 20\% and for SNR by 200\% (see main text). The hatched region indicates where no signal was recorded. Blue circle and red diamond indicate trajectories shown in (a). (d)~Theoretical phase diagram showing qualitative agreement with the experimental results. The hatched region indicates the parameter region below condensation threshold with $N_1=N_2=0$.}\label{fig:3}
\end{figure*}

\subsection*{Observation of Limit Cycles}
We first demonstrate the presence of limit cycle dynamics in the experiment. As our key observables, we measure the photon numbers $N_{1,2}$ in a site-resolved fashion via two fast photodetectors with $\SI{0.52}{\nano\second}$ temporal resolution. Figure~\ref{fig:2}(a) shows a representative single-shot measurement of the emission dynamics at constant pump power. Note that the measured photon number signals are normalised by their respective mean values. We start the measurement with a detuned double-well potential, as realised by a weakly-tilted cavity geometry, and observe a steady-state emission in both $N_1$ and $N_2$. As time evolves, three distinct dynamical regimes are identified [see Fig.~\ref{fig:2}(b)], which result from a slow thermal drift of the frequency detuning $\Delta=\Delta_1-\Delta_2$ between the two sites (see Methods). The drift is attributed to a thermo-optic refractive index change by laser-induced heating of the dye medium at the pumped site~\cite{Dung:2017,Schmitt:2018}, leading to an approximately linear shift of its resonance frequency relative to the unpumped site, as illustrated in the top of Fig.~\ref{fig:2}(a). Since limit cycles occur only close to resonance between the two sites, we use the drift to effectively sweep the system through the oscillatory regime. As the detuning drift rate of about $\SI{2}{\mega\hertz\per\nano\second}$ is orders of magnitude slower than the cavity loss rate $\kappa\approx \SI{6}{\giga\hertz}$, it can be considered adiabatic. 

Figure~\ref{fig:2}(b) shows three time windows illustrating the system's evolution. Before reaching resonance $\Delta>0$ (left panel), the system is in a stationary state, with the emitted photon numbers relaxing to constants. The corresponding ($N_1$-$N_2$)-histogram in Fig.~\ref{fig:2}(c) is concentrated near a single point, consistent with a single steady state. Near resonance $\Delta=0$ (middle panels), the photon numbers exhibit persistent oscillations in time with $\nu_{\mathrm{LC},1}=\SI{283\pm 3}{\mega\hertz}$ and $\nu_{\mathrm{LC},2}=\SI{281\pm 2}{\mega\hertz}$. The histogram forms a ring-shaped distribution, reflecting the closed orbit associated with a limit cycle. After the resonance is crossed $\Delta<0$ (right panels), the system returns to a stationary state indicated by the absence of oscillations and a histogram again concentrated on a single point. The observation of self-sustained nonlinear dynamics with $\nu_\mathrm{LC}\ll J$ provides a first line of evidence for reservoir-mediated dissipative interactions between photons in the double-well system.


\subsection*{Dissipative Phase Diagram}
Next, we show the tuneability of the effective interaction strength $\mathcal{G}_j$ according to Eq.~\eqref{eq:interaction}, by investigating the appearance of limit cycles as a function of the experimental pump rate and cutoff wavelength. For this, we analyse the temporal and spectral properties of the emission for the on-resonant case $\Delta=0$. Figure~\ref{fig:3}(a) shows time traces of the normalised photon number in the pumped well, $N_1/\bar{N}_1$, for two different pump powers and at a cutoff wavelength $\lambda_\mathrm{c}=\SI{588.5}{\nano\meter}$. At the lower pump power $P_1=\SI{2.78}{\micro\watt}$ (top), the system is prepared in a steady state which only exhibits fluctuations around a constant mean. At the larger power $P_1=\SI{4.44}{\micro\watt}$ (bottom), we encounter a regime with persistent oscillations. The peak in the power spectrum $S(\nu)$ in Fig.~\ref{fig:3}(b) at frequency $\nu_\mathrm{LC}$ and spectral width $\Delta\nu$ provides a clear spectral signature of the limit cycle. In contrast, the stationary regime exhibits only broadband noise without a distinct frequency component. 

We use these spectral signatures to systematically map out the occurrence of limit cycles across the parameter space of $\lambda_
\mathrm{c}$ and $P_1$, resembling a phase diagram of the nonequilibrium dynamics. Pump powers ranging from $P_1=\SI{1.4}{\micro\watt}$ to $\SI{4.9}{\micro\watt}$ and cutoff wavelengths from $\lambda_\mathrm{c}=\SI{585}{\nano\meter}$ to $\SI{595}{\nano\meter}$ were investigated. Figure~\ref{fig:3}(c) shows the average power spectral density at the frequency $\nu_\mathrm{LC}$ (top right panel), the spectral width $\Delta\nu$ (bottom right), and the probability $\mathcal{P}_\mathrm{LC}$ of observing limit cycles -- calculated as the fraction of trajectories with $\Delta\nu<\SI{20}{\mega\hertz}$ and a signal-to-noise ratio $>100$ at $\nu_\mathrm{LC}$ -- versus $\lambda_\mathrm{c}$ and $P_1$ (main panel). The parameter range with a clear appearance of a spectral peak $S(\nu_\mathrm{LC})$, together with a decrease of $\Delta\nu$, identifies the regime where stable limit cycles are present, visible as a blue region in the main panel. In contrast, red regions indicate a regime where the system is governed by a stable fixed point. The extent of the oscillatory region depends on both $P_1$ and $\lambda_\mathrm{c}$, demonstrating that the non-Hermitian interaction can not only stabilise limit cycles but is also tuneable in strength. We attribute false negatives within the LC-region (\textit{e.g.}, red data at $\lambda_\mathrm{c}\approx\SI{590}{\nano\meter}$ and $P_1\approx \SI{4}{\micro\watt}$) and false positives within the FP-region (\textit{e.g.}, blue data at $\lambda_\mathrm{c}\approx\SI{595}{\nano\meter}$ and $P_1\approx \SI{1.5}{\micro\watt}$) to arise from transient oscillations when the double-well detuning $\Delta$, the cavity cutoff $\lambda_\mathrm{c}$ or the pump power $P_1$ have drifted between different measurement runs due to technical variations in the experimental apparatus.

To validate the phase diagram, we have performed a numerical stability analysis [as in Fig.~\ref{fig:1}(d)] as a function of the experimental control parameters. Figure~\ref{fig:3}(d) shows the resulting parametrically unstable region in blue, forming an island in parameter space where the stationary state undergoes a Hopf bifurcation and limit-cycle dynamics emerge. Residual discrepancies between theory and experiment may arise from limited accuracy for the cavity loss rates, which have not been measured for the used double-well structured mirrors. The overall agreement between experiment and theory demonstrates that the observed dynamics are well captured by our model and supports their interpretation in terms of effective dissipative photon interactions.


\subsection*{Hysteresis and Bistability}
Finally, we show that purely dissipative interactions exhibit hysteresis and bistable behaviour, akin to systems with conventional Kerr nonlinearities. Unfortunately, for large $\Gamma_1^\uparrow$ and reduced $\lambda_\mathrm{c}$, the expected frequencies between $\nu_\mathrm{LC}\approx \SI{1.5}{\giga\hertz}$ and $\SI{5}{\giga\hertz}$ are too large to be resolved temporally, yet too small to be resolved spectrally in the present experiment. We therefore exploit the good agreement between experiment and numerics of Eq.~\eqref{eq:model} (Methods) to extend our analysis with stochastic simulations into the relevant parameter regime. 

\begin{figure}[t]
    \includegraphics[width=\columnwidth]{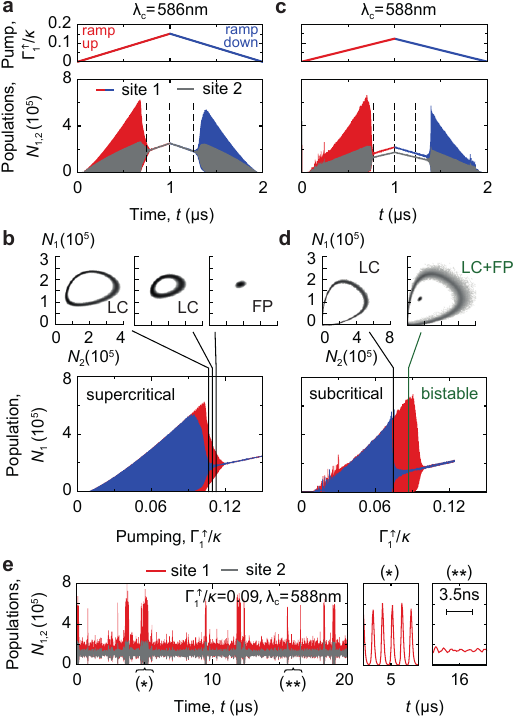}
    \caption{\textbf{Super- and subcritical dynamics.} (a) Simulated photon numbers $N_{1,2}$ at site 1 and 2 (bottom) for $\lambda_\mathrm{c}=\SI{586}{\nano\meter}$ as the pump rate $\Gamma_1^\uparrow$ is swept linearly in time (top), while $\Gamma_2^\uparrow$=0. Note that the limit cycle oscillations are not resolved in time and only the amplitude is visible as red, blue and grey regions. (b) $N_1$ versus pump. Weak hysteresis is observed between up (red) and down (blue) ramps. The histograms obtained from stochastic simulations at fixed pump rates show only a single steady state or limit cycle (top). (c) $N_{1,2}$ as a function of time, now for a different cutoff wavelength $\lambda_\mathrm{c}=\SI{588}{\nano\meter}$ and (d) $N_1$ versus pump, showing pronounced hysteresis. Histograms at a fixed pump values show only a limit cycle (top left) or coexistence of a central stationary state and a limit cycle (top right). (e) Stochastic time trace in the bistable regime, showing noise-activated transitions between stationary and oscillatory dynamics. Panels on the right give a zoomed view of indicated temporal windows.}
     \label{fig:4}
\end{figure}

Figure~\ref{fig:4}(a) shows the response of $N_{1,2}$ under a slow up-and-down ramp of the pump rate $\Gamma_1^\uparrow$ at $\lambda_\mathrm{c}=\SI{586}{\nano\meter}$. Weak hysteresis is observed between the limit cycles as a function of $\Gamma^\uparrow_1$, as shown in Fig.~\ref{fig:4}(b). Histograms at $\Gamma^\uparrow_1$ within the hysteresis window show either only limit cycles, or only a stable steady state (top), indicating that the hysteresis originates from the finite reservoir response time. Such non-adiabatic behaviour is distinct from hysteresis associated with bistability and does not imply the coexistence of multiple stable states. In fact, it is well consistent with a supercritical Hopf bifurcation, for which limit cycles emerge continuously and no such coexistence is expected. Strikingly, at the nearby $\lambda_\mathrm{c}=\SI{588}{\nano\meter}$, bistability emerges. Figures~\ref{fig:4}(c,d) show the clear hysteresis when scanning $\Gamma^\uparrow_1$. The simultaneous occurrence of a stationary state peak and a ring-shaped limit cycle distribution in the histogram confirms the coexistence of the two attractors. Moreover, noise-activated switching events between the stationary fixed point and the limit cycle in the bistable regime are found and visible in Fig.~\ref{fig:4}(e). The combined findings in Figs.~\ref{fig:4}(c-e) are characteristic of subcritical Hopf bifurcations and a fold of cycles --- a bifurcation structure with rich excitable and bistable dynamics found in many biological systems~\cite{Izhikevich:2000}. The observed dependence of the dynamical behaviour on $\lambda_\mathrm{c}$ reflects the sensitivity of the Hopf bifurcation to the absorption and emission rates, which control both linear gain and nonlinear dissipative interaction.

\section*{Discussion}
To conclude, we have demonstrated tuneable dissipative interactions for light. The nonlinearity is directly revealed by the observation of limit-cycle dynamics in two photon condensates in a dye-filled double-well microcavity. Physically, the self-sustained oscillations originate from the interplay between coherent photon tunnelling and local reservoir-induced gain imbalance and saturation. By tuning the effective interaction strength, a phase diagram with domains where either limit cycles or stable fixed points occur is found. The nature of the phase boundary is inferred from simulations, which reveal super- and subcritical Hopf bifurcations, with the latter showing hysteresis behaviour. In the supercritical regime, a limit cycle acts as the sole attractor in phase space, while in the subcritical regime, stationary and oscillatory states coexist. This demonstrates that attractor landscapes for many-body systems can be engineered solely by dissipative interactions.  

Beyond the present work, the class-II excitability numerically predicted for our system, characterised by a finite-frequency oscillatory onset at $\mathcal{G}_j<0$, provides key ingredients for neuromorphic architectures with deterministic switching capabilities~\cite{Izhikevich:2000,Markovic:2020}. Combined with the recent demonstrations of photon Bose-Einstein condensation in semiconductor microcavities~\cite{Schofield:2024,Pieczarka:2024}, these results open a route toward a novel class of physical computation devices based on reservoir-mediated dynamics. More broadly, our results bring forward a versatile testbed for exploring quantum--classical crossovers in limit cycles~\cite{Ben:2021} and stochastic thermodynamics, including entropy production and fluctuation relations~\cite{Falasco:2025,Santolin:2025}.


\setcounter{figure}{0}
\makeatletter
\renewcommand{\fnum@figure}{\textbf{Supplementary Fig.~\thefigure~|}}
\renewcommand{\@caption@fignum@sep}{\enspace}
\makeatother
\renewcommand{\theHfigure}{Methods.\thefigure}

\section{Methods}

\subsection{Experimental methods}
Our experiment is performed in a microstructured optical microcavity filled with a liquid dye solution. The used Rhodamine 6G dye is solved in ethylene glycol (refractive index $\tilde n\approx 1.44$) at a concentration of $\SI{1}{\milli\mol\per\liter}$~\cite{Schmitt:2024}. The cavity is formed by two Bragg mirrors with reflectivity $R \approx 99.991(3)\%$ spaced by $D_0\approx\SI{1.4}{\micro\meter}$~\cite{Schmitt:2018}). The free spectral range in the resonator $\Delta\nu=c/(2\tilde n D_0) \approx \SI{73}{\tera\hertz}$ exceeds the emission bandwidth of the dye molecules (of order $k_\mathrm{B}T/h\approx\SI{6.2}{\tera\hertz}$), so that the photons are emitted only into modes with a fixed longitudinal mode number $q=7$, introducing a low-frequency cutoff wavelength $\lambda_\mathrm{c}={2\tilde n D_0}/{q}\approx\SI{585}{\nano\meter}$. Note that the actual dye film thickness between the cavity mirrors is roughly $\SI{0.47}{\micro\meter}$ due to the penetration of the optical field into the dielectric mirrors. For this dye film thickness, one expects $M\approx 5\cdot 10^6$ dye molecules contained within the single-well mode volume of waist $w_0\approx \SI{3.5}{\micro\meter}$ at the given concentration, which closely agrees with the experimentally observed molecule numbers. The photon dispersion relation in the paraxial approximation ($k_{x,y}\ll k_z$) is equivalent to that of a two-dimensional massive Bose gas,
\begin{equation}
	E \simeq \frac{\hbar^2 (k_x^2+k_y^2)}{2m_\mathrm{ph} }+ m_\mathrm{ph} \left(\frac{c}{\tilde n}\right)^2 \left(1+\frac{\Delta D}{D_0} + \frac{\Delta \tilde n}{\tilde n}\right),
	\label{eq:dispersion}
\end{equation}
where $m_\mathrm{ph}=2\pi\hbar n^2/(\lambda_\mathrm{c}c)=\SI{7.8e-36}{\kilogram}$ denotes the effective photon mass, $c$ the vacuum speed of light, and $k_{x,y}$ the transverse components of the wave vector $\mathbf{k}=(k_x,k_y, \pi /\tilde n D_0)$.

By introducing a local cavity length variation $\Delta D(x,y)$, we control the energy landscape for the photons and realise a double-well trap potential. For this, the surface of one of the mirrors is elevated except for two concave indents, as shown in Fig.~\ref{fig:1}(b) of the main text, using direct laser writing~\cite{Kurtscheid:2020,Vretenar:2023}. The separation between the minima of the indents is $\SI{10.9\pm0.1}{\micro\meter}$, and the structure depth is $\SI{1.4\pm0.1}{\nano\meter}$. In the cavity, $\Delta D(x,y)$ corresponds to a trap depth $U_\mathrm{D}/\hbar=\hbar^{-1}m_\mathrm{ph} (c/\tilde n)^2\,\Delta D/D_0=2\pi\times \SI{504\pm36}{\giga\hertz}$ (in frequency units). We tune both $\lambda_\mathrm{c}$ (via cavity length) and the detuning of the double-well (via cavity tilt) using piezoelectric nanopositioners integrated within the microcavity apparatus; as indicated in Fig.~\ref{fig:2}(a), the experiment is initiated with the pumped site slightly red-detuned with respect to the unpumped site.

To excite the photonic dimer in the double well the dye solution is pumped with a laser beam at $\SI{532}{\nano\meter}$ wavelength, which is acousto-optically chopped into $\SI{3}{\micro\second}$-long pulses at $
\SI{10}{\hertz}$ to minimise triplet-state excitation of the dye--molecules. Depending on the pump beam intensity profile, the dye medium can provide either spatially-varying or uniform gain; the former prepares the system further away from equilibrium. The present work studies the nonequilibrium dynamics emerging under imbalanced pumping, by exciting molecules at a single site of the double well, while the second site remains weakly pumped. The experimentally used pump ratio $\Gamma_1^\uparrow/\Gamma_2^\uparrow$ is roughly $9$:$1$.

The photons in the double-well system are analysed by imaging the cavity emission from the left and right well, respectively, on two photomultipliers with a temporal resolution of $\SI{0.52}{\nano\second}$. The time-resolved output voltage time traces are sampled by an oscilloscope with $\SI{6}{\giga\hertz}$ bandwidth and recorded in single-shot measurements for repeated pump pulses. By normalizing the data, we obtain the evolution of the relative photon numbers $N_1(t)/\bar N_1$ and $N_2(t)/\bar N_2$. Power spectral densities are obtained from Fourier transforms of the measured time traces. Histograms in the $(N_1,N_2)$ plane are constructed from ensembles of repeated trajectories acquired under the same control parameters.

\subsection{Measurement of coherent intracavity coupling $J$}

The coupling rate $J$ of the photonic dimer in the double well is determined by a time-resolved measurement of the coherent Rabi dynamics in the effective two-level system. Note that purely coherent dynamics is distinct from the driven-dissipative limit-cycle dynamics reported in the main text and requires an alternative excitation scheme. For this, dye molecules at one well are excited using a focused mode-locked laser beam with $\SI{20}{\pico\second}$ pulse duration. The temporal evolution of the photon numbers in the pumped and unpumped well, respectively, is modelled by
\begin{subequations} \label{eq:system}
	\begin{align}
		N_1&=A\, e^{-\varkappa t}\left[\cos^2\left(\Omega t-\phi\right)+\frac{\Delta^2}{4\Omega^2}\sin^2\left(\Omega t-\phi\right)\right]     \label{eq:fitmodel1} \\
		N_2&=A\, e^{-\varkappa t}\frac{J^2}{\Omega^2}\sin^2\left(\Omega t-\phi\right), \label{eq:fitmodel2}
	\end{align}
\end{subequations}
where $\varkappa$ denotes a phenomenological loss, $A$ an amplitude, $\phi$ a phase, $\Omega=\sqrt{J^2+{\Delta^2}/{4}}$ the effective Rabi frequency and $\Delta=\Delta_1-\Delta_2$ the detuning, see Eq.~\eqref{eq:model} of the main text. The coherent intracavity coupling $J$ is obtained from the effective Rabi frequency $\Omega$ and the detuning $\Delta$ for near resonant conditions $\Delta=0$.

\begin{figure}[t]
	\includegraphics[width=\columnwidth]{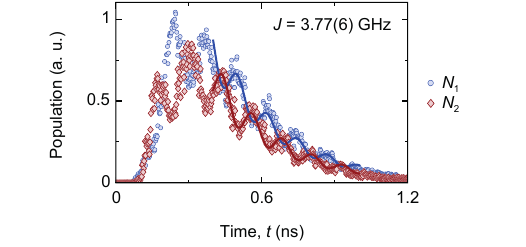}
	\caption{\textbf{Coherent photon dynamics.} Time-resolved population $N_{1}$ and $N_2$ of the pumped (blue circles) and unpumped site (red diamonds), respectively, in the double well potential, along with fits (solid lines). The double well structure corresponds to the one shown in Fig.~\ref{fig:1}(b) of the main text. After a pulsed excitation of the molecules at time $t\approx \SI{100}{\pico\second}$ times, the photon population grows and exhibits coherent oscillations between both sites, followed by a decrease caused by photon losses from cavity mirror transmission.} \label{fig:S1}
\end{figure}

To achieve a spatially localised pulsed excitation at one double well site, we first record the photon density at large detuning $|\Delta|\gg J$, so that the mixing of the eigenmodes is weak. In this way, we ensure that the molecule emission predominantly occurs only into the (decoupled) mode that overlaps with the localised pump spot. We actively control $\Delta$ by tilting one of the cavity mirrors. Before starting the time-resolved measurement, we again set $\Delta\approx 0$ by minimizing the mirror tilt to achieve equal photon populations at both sites in position space and a balanced mixing of the symmetric and antisymmetric eigenstate probability densities, as verified by a grating spectrometer with spatial resolution. As the eigenstate mixing depends on $\Delta/J$, see Eq.~\eqref{eq:model} of the main text, this allows us to realise $\Delta\approx 0$.

Extended Data Fig.~\ref{fig:S1} shows the measured population oscillations in a double well with $|\Delta| < J$. The data was recorded by repeatedly initializing the system with the excitation pulse and subsequently recording the position-resolved cavity emission $I(x,y)$ on an ICCD camera with a temporal resolution of $\SI{69\pm3}{\pico\second}$ and a variable time delay $\tau$. As the recorded signal is a convolution of the photon number dynamics and the detector response, the contrast of the oscillations is limited. We fit the data with Eqns.~\eqref{eq:fitmodel1} and \eqref{eq:fitmodel2}, yielding the tunnel coupling $J=2\pi\times\SI{3.77 \pm 0.06}{\giga\hertz}$, detuning   $|\Delta| = 2\pi\times\SI{2.93 \pm 0.16}{\giga\hertz}$, and apparent cavity loss $\varkappa =\SI{3.60 \pm 0.09}{\giga\hertz}$. Averaging over many repetitions, we obtain a tunnel coupling $J=2\pi\times\SI{3.6\pm 0.1}{\giga\hertz}$, as reported in the main text. Note that to exclude the build-up dynamics at early times (which is not captured by the fit model) and to minimise the influence of cavity drifts during the measurement, we limit our fit to the $\SI{0.5}{\nano\second}$ to $\SI{1.1}{\nano\second}$ time range shown in Extended Data Fig.~\ref{fig:S1}. The fit value for $\varkappa$ should be taken with care, because the fit model assumes only photon losses through the cavity mirrors, although absorption and emission processes and molecule decay modify the loss rate. Furthermore, the losses depend on wavelength; for the experiment in Extended Data Fig.~\ref{fig:S1}, $\lambda_c=\SI{583.5}{\nano\meter}$, while in the main text ranges from $\lambda_c=\SI{584.5}{\nano\meter}$ to $\lambda_c=\SI{597.5}{\nano\meter}$. According to the mirror reflectivity, the actual cavity loss rate $\kappa$ is estimated to be $\kappa^\mathrm{(exp)}\approx\SI{5.7}{\giga\hertz}$ ($\lambda_c=\SI{584.5}{\nano\meter}$) and $\kappa^\mathrm{(exp)}\approx\SI{24.3}{\giga\hertz}$ ($\lambda_c=\SI{597.5}{\nano\meter}$). However, as the fit results for $J$ and $\Delta$ are independent of $\varkappa$, the measurements provide a reliable extraction of the $J$ parameter.

\subsection{Theoretical Model}

In this section we derive the stochastic model used in the main text. We start from the Hamiltonian and Lindblad master equation, then move to a phase-space formulation to obtain a Fokker-Planck equation for the Wigner function, which directly maps to a stochastic differential equation (SDE) for the cavity fields. For simplicity we start from a single site, from which the extension to coupled sites is trivial. 

Our system comprises a dye-filled optical microcavity. The optical mode volume contains $M$ dye molecules. In a frame rotating at the molecule zero-phonon transition frequency $\omega_\mathrm{zpl}$, the Hamiltonian is given by
\begin{equation}
	\hat{\mathcal{H}} = \hbar\Delta\hat{a}^\dagger \hat{a},
\end{equation}
with photon annihilation (creation) operators $\hat{a}^{(\dagger)}$ and detuning $\Delta=\omega-\omega_\mathrm{zpl}$ of the cavity cutoff frequency $\omega$. The Jaynes-Cummings interaction that couples photons to the electronic molecular transitions is omitted because fast dephasing renders coherent exchange overdamped, yielding an effective incoherent rate description. The molecules therefore act as an incoherent particle reservoir for the photons.

The Hamiltonian accounts only for coherent processes. To include dissipative processes, we employ the Lindblad formalism using jump operators $\mathcal{D}[\hat F](\hat{\rho})=\hat F\hat{\rho} \hat F^\dagger-\tfrac12\{\hat F^\dagger \hat F,\hat{\rho}\}$ of some operator $\hat F$ and density operator $\hat \rho$. In our system dissipation arises from cavity photon loss at rate $\kappa$, molecular absorption and emission at rates $B_\textrm{abs}$ and $B_\textrm{em}$, respectively, incoherent pumping at rate $\Gamma^\uparrow$, and nonradiative decay at rate $\Gamma^\downarrow$. The Lindblad master equation for the driven-dissipative photon-molecule system~\cite{Kirton:2013,Abouelela:2025} is then
\begin{equation}\label{eq:Lindblad}
	\begin{aligned}        
		\partial_t\hat\rho  = &  -\frac{i}{\hbar}[\hat{\mathcal{H}},\hat\rho]
		+ \kappa \mathcal{D}\left[\hat{a}\right]\hat{\rho} \\
		& + \sum_{n=1}^M (B_\textrm{abs}\mathcal{D}\left[\hat{a}\hat{\sigma}_{n}^+\right]\hat{\rho}+B_\textrm{em}\mathcal{D}\left[\hat{a}^\dagger\hat{\sigma}_{n}^-\right]\hat{\rho} \\
		&\ \ \ \ \ \ \ \ \  + \Gamma^\uparrow\mathcal{D}\left[\sigma_{n}^+\right]\hat{\rho} + \Gamma^\downarrow\mathcal{D}\left[\sigma_{n}^-\right]\hat{\rho} ).
	\end{aligned}
\end{equation}

To derive stochastic field equations we transform the master equation to Wigner phase space. We use the Wigner correspondences for a bosonic mode,
\begin{equation}
	\begin{aligned}
		\hat a\,\hat\rho &\;\longrightarrow\; \left( \psi + \frac12 \partial_{\psi^*}\right)W(\psi,\psi^*,t),\\
		\hat\rho\,\hat a &\;\longrightarrow\; \left( \psi - \frac12 \partial_{\psi^*}\right)W(\psi,\psi^*,t),\\
		\hat a^\dagger\hat\rho &\;\longrightarrow\; \left( \psi^* - \frac12 \partial_{\psi}\right)W(\psi,\psi^*,t),\\
		\hat\rho\,\hat a^\dagger &\;\longrightarrow\; \left( \psi^* + \frac12 \partial_{\psi}\right)W(\psi,\psi^*,t),
	\end{aligned}
\end{equation}
where $W(\psi,\psi^*,t)$ is the Wigner function associated with $\hat\rho(t)$ and 
$\psi,\psi^*$ are complex phase-space variables. 
Ensemble averages over $W$ reproduce symmetrically ordered operator moments; in particular
$\langle \hat a\rangle = \langle \psi\rangle_W$ and $\langle \hat a^\dagger\rangle = \langle \psi^*\rangle_W$. Applying this transformation leads to the following Wigner-Fokker-Planck equation
\begin{widetext}
	\begin{equation}\label{eq:FP}
		\begin{aligned}
			\frac{\partial W}{\partial t}
			=&
			i\Delta\left[\partial_\psi ( \psi W) - \partial_{\psi^*}(\psi^* W)\right]
			+
			\frac12 \left[\kappa + B_\textrm{abs} M (1-m) - B_\textrm{em} M m\right]
			\left[
			\partial_\psi(\psi W) +\partial_{\psi^*}(\psi^* W)
			\right] \\
			&+
			\frac12 \left[\kappa + B_\textrm{abs} M (1-m) + B_\textrm{em} M m\right]
			\partial_\psi \partial_{\psi^*}W,
		\end{aligned}
	\end{equation}
\end{widetext}
where we defined the excited molecule fraction
\begin{equation}\label{eq:molfrac}
	m = \frac{1}{M}\sum_{n=1}^M\left\langle \hat\sigma_n^+\hat\sigma_n^- \right\rangle.
\end{equation}
Equation~\eqref{eq:FP} is of the form
\begin{equation}
	\frac{\partial W}{\partial t} = -\sum_i \partial_{x_i}(A_iW) + \sum_{ij}\partial_{x_i}\partial_{x_j}(D_{ij}W),
\end{equation}
with $x_i\in (\psi,\psi^*)$, drift terms $A_i$ and diffusion $D_{ij}$, such that it directly maps to an SDE for the field:
\begin{equation}\label{eq:SDE}
	\begin{aligned}
		\dot{\psi} &= A(\psi,m) + \sqrt{D(m)}\;\xi(t),\\
		A(\psi,m) &= \left[-i\Delta + \frac12 \left[-\kappa - B_\textrm{abs}M\left(1-m\right) + B_\textrm{em}M m\right]\right]\psi\\
		D(m) &= \frac12 \left[\kappa + B_\textrm{abs}M\left(1-m\right) + B_\textrm{em}M m\right].
	\end{aligned}
\end{equation}
Here, $\xi(t)$ is a complex Gaussian process with $\left\langle \xi(t)\right\rangle=0$ and $\left\langle \xi(t)\xi^*(t')\right\rangle=\delta(t-t')$.

For our experiments, the molecule number $M\gg1$, so that fluctuations in the fraction of excited molecules can be neglected, as they scale with $1/\sqrt{M}$. The equation of motion for the excited molecule fraction $m$ can then be obtained by using Eqns.~\eqref{eq:Lindblad} and \eqref{eq:molfrac} together with $\partial_t \langle \hat F\rangle=\mathrm{Tr}[ \hat F\partial_t\hat\rho]$. Assuming correlation functions factorise as $\langle \hat F_1\hat F_2\rangle=\langle \hat F_1\rangle\langle \hat F_2\rangle$, we arrive at
\begin{equation}\label{eq:mol_ode}
	\begin{aligned}
		\dot{m}=\ &\Gamma^\uparrow \left( 1-m \right) - \Gamma^\downarrow m \\
		&+ B_\textrm{abs}(1-m)\left(|\psi|^2-\frac12 \right) - B_\textrm{em} m \left(|\psi|^2+\frac12\right),
	\end{aligned}
\end{equation}
where we used $\langle \hat a^\dagger\hat a\rangle = \langle |\psi|^2-\frac12\rangle_W$ and $\langle \hat a\hat a^\dagger\rangle= \langle |\psi|^2+\frac12\rangle_W$. Finally, extending the field equation to two sites coupled coherently with rate $J$, neglecting the constant $\pm\frac12$ corrections since $|\psi|^2\gg 1$ for our parameters, and multiplying by $i\hbar$ yields Eq.~\eqref{eq:model} of the main text.

\subsection{Simulation details}
Our stochastic simulations are based on the xSPDE toolbox for  Matlab~\cite{xSPDE}. We employ a fourth-order Runge-Kutta algorithm, and set a time increment $\Delta t=5$~ps, which is much smaller than $\kappa^{-1}=\mathcal{O}(\SI{1}{\nano\second})$, for all parameter regimes that are simulated.

\begin{figure}[!t]
	\includegraphics[width=\columnwidth]{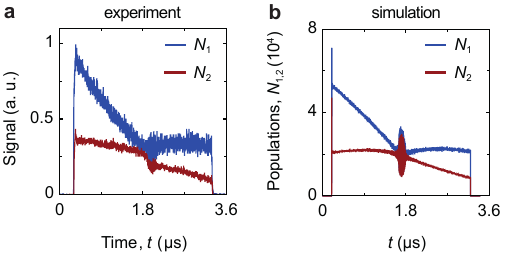}
	\caption{\textbf{Comparison of experiment with simulation.} (a)~Full measured photon number trajectories $N_{1,2}$ in the pumped and unpumped well, respectively, for $\lambda_\mathrm{c}=\SI{588}{\nano\meter}$ and $\SI{3.42}{\micro\watt}$. (b)~Simulated trajectories after convolution with a $\SI{0.52}{\nano\second}$ impulse response of the used detector. Simulation parameters: $\lambda_c=\SI{588}{\nano\meter}$, $\Gamma_1^\uparrow=2\pi\times\SI{200}{\mega\hertz}$, $\Gamma_2^\uparrow=\Gamma_1^\uparrow/10$, $\Gamma^\downarrow=2\pi\times\SI{250}{\mega\hertz}$, $M=4.4\times 10^6$, $J=2\pi\times\SI{3.6}{\giga\hertz}$, $\gamma_\mathrm{d}=\SI{700}{\kilo\hertz}$, and $v_\Delta=\SI{2}{\mega\hertz\per\nano\second}$.} \label{fig:S2}
\end{figure}

\subsection{Comparison of experiment and simulated traces}

We compare the experimental and numerically calculated photon number time traces. Extended Data Fig.~\ref{fig:S2}(a) shows the emitted intensities during the full quasi-CW pump pulse experiment. Molecular depletion by the pump results in an overall decay of the emission over time. Additionally, pump-induced local heating results in a slow drift of the inter-well detuning, and consequently a change in the imbalance over time.

To replicate these experimental conditions, we defined a square-wave temporal profile for $\Gamma^\uparrow_1$ (with `on'-time $T$) to reflect the experimental quasi-CW pump. Additionally, molecular depletion is modelled by an additional differential equation for the number of molecules
\begin{equation}
	\dot{M}_j=-\gamma_\mathrm{d} M_j m_j
\end{equation}
where $\gamma_\mathrm{d}$ is the molecular depletion rate. The slow detuning drift is approximated by a linear time dependence of the detuning of the pumped site,
\begin{equation}
	\Delta_1(t)=v_\Delta\left(t-\frac{T}{2}\right),
\end{equation}
with drift rate $v_\Delta$, while $\Delta_2(t)=0$. To compare experiment and theory on equal footing, the finite temporal resolution of the detector was taken into account in the simulated photon number traces by convolving the data with an impulse response corresponding to the measured detector rise time of $\SI{0.52}{\nano\second}$. This suppressed frequency components above the effective detector bandwidth and reduces the oscillation amplitudes. Extended Data Fig.~\ref{fig:S2}(b) shows the resulting intracavity photon numbers obtained from numerical simulations, which agree with the experimental observations for $\gamma_\mathrm{d}=\SI{700}{\kilo\hertz}$ and $v_\Delta=\SI{2}{\mega\hertz\per\nano\second}$.

Near $\Delta_1(T/2) \approx 0$, both traces exhibit self-sustained oscillations, which we identify as limit cycles in Fig.~\ref{fig:2} of the main text. Intuitively, the limit cycles can be understood to emerge from the interplay between coherent transport and effective interactions: The coherent coupling between the two sites drives population oscillations, while differences in the effective gain and loss rates damp this exchange. When the coherent coupling overcomes the dissipative imbalance, the modes acquire an oscillatory character and grow until reservoir-mediated gain saturation sets in and stabilises a nonequilibrium limit cycle.

\subsection{Effective non-Hermitian interactions}
To obtain an analytical expression for the effective photon-photon interactions arising from the coupling to dye molecules, we make use of the adiabatic approximation. On the timescale of the slow collective photon-molecule dynamics near the Hopf bifurcation, the molecular excitation fraction can be approximated as quasi-stationary. We set $\dot m\simeq 0$ and solve Eq.~\eqref{eq:mol_ode} algebraically for $m$ as a function of intensity $N=|\psi|^2$. Expanding for small $N$ gives
\begin{equation}\label{eq:mol_expansion}
	m = m^{(0)} + \chi\, N + \mathcal{O}(N^2),
\end{equation}
with
\begin{align}
	m^{(0)} &= \frac{\Gamma^\uparrow}{\Gamma^\uparrow+\Gamma^\downarrow} , \\
	\chi &= \frac{  B_\textrm{abs}\Gamma^\downarrow - \Gamma^\uparrow B_\textrm{em} }
	{(\Gamma^\uparrow+\Gamma^\downarrow)^2}.
\end{align}
Substituting Eq.~\eqref{eq:mol_expansion} into Eq.~\eqref{eq:SDE} yields 
\begin{equation}\label{eq:adiabatic_elim}
	i\hbar\dot\psi = \left[\hbar\Delta + i\hbar G^{(0)} + i\hbar  
	\mathcal{G}|\psi|^2\right]\psi
\end{equation}
with effective linear gain coefficient
\begin{equation}
	G^{(0)} = \frac12\left(-\kappa -B_\textrm{abs} M + \left( B_\textrm{abs} + B_\textrm{em} \right) M m^{(0)}\right)
\end{equation}
and non-Hermitian interaction strength
\begin{equation}
	\mathcal{G}=\frac12\left(B_\textrm{abs}+B_\textrm{em}\right) M \chi.
\end{equation}
Note that both the linear and nonlinear terms, governed by $G^{(0)}$ and $\mathcal{G}$, respectively, are acting locally on the field $\psi$. In the double-well system of two coupled cavity modes $\psi_j$ ($j=1,2$) studied in the present work, the terms thus become effectively site-dependent due to the locally varying pump excitations $\Gamma_j^{\uparrow}$.

The nonlinear term $i\hbar \mathcal G |\psi|^2\psi$ can formally be interpreted as an imaginary Gross--Pitaevskii-type interaction. In analogy to the dimensionless Hermitian interaction parameter $\tilde g$ of two-dimensional Bose gases~\cite{Klaers:2010}, this corresponds to an effective dissipative coupling $\tilde g_\mathrm{diss}\propto i m_\mathrm{ph}\mathcal G/\hbar$. We note, however, that the present system realises a driven-dissipative photonic dimer rather than an extended two-dimensional condensate, such that this correspondence is formal and depends on the specific field normalisation. The present platform nevertheless provides experimental access to a regime of purely non-Hermitian many-body interactions that is fundamentally distinct from conservative thermo-optic or Kerr nonlinearities.

\subsection{Validity of the effective interaction picture}
At first sight, adiabatic elimination of the molecular reservoir may appear questionable, since the bare cavity lifetime is shorter than the molecular relaxation time, approximately given by
\begin{equation}
	\Gamma_\mathrm{res}\sim\Gamma^\uparrow+\Gamma^\downarrow+\left(B_\mathrm{abs}+B_\mathrm{em}\right)|\psi|^2.
\end{equation}
For typical parameters in this study $\Gamma_\mathrm{res}$ and $\kappa$ are of the same order of magnitude, which would prevent adiabatic elimination of the molecular reservoir.

However, the relevant comparison is not between the reservoir dynamics and the bare cavity lifetime, but between the reservoir relaxation and the slow collective dynamics near the instability. Close to the Hopf bifurcation, the relevant order-parameter mode becomes critically slowed down, such that the real part of the corresponding linear stability eigenvalue approaches zero. The observed limit-cycle dynamics therefore occur on timescales slower than the reservoir relaxation. On this coarse-grained timescale, the molecular excitation fraction follows the slowly varying condensate intensity quasi-statically, which justifies approximating the reservoir by its instantaneous stationary value and setting $\dot m\simeq0$.

\begin{figure}[t]
	\includegraphics[width=\columnwidth]{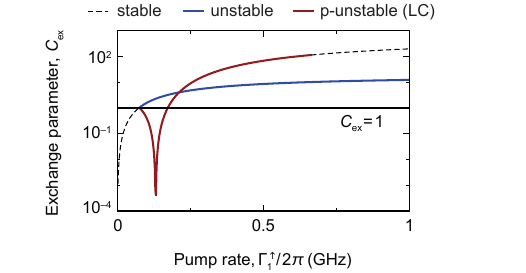}
	\caption{\textbf{Effective interaction picture.} Calculated exchange parameter $C_\mathrm{ex}$ as function of the pump rate $\Gamma_1^\uparrow$ at the first double well site within the adiabatically eliminated theory. Stability of the branches is computed using the full photon--reservoir model from Eq.~\eqref{eq:model}. Black dashed curve: stable fixed point; blue solid curve: single-mode unstable; red solid curve: parametrically unstable. Horizontal line indicates $C_\mathrm{ex}=1$. Parameters are as in Fig.~\ref{fig:1}(d) of the main text.} \label{fig:S3}
\end{figure}

The resulting effective interaction picture can be tested by comparing the oscillatory instability obtained from the full photon--reservoir model Eq.~\eqref{eq:model} with the prediction of the adiabatically eliminated theory. Linearising Eq.~\eqref{eq:adiabatic_elim} around the stationary photon numbers $N_j^{(s)}$ yields the effective local gain coefficients
\begin{equation}
	\alpha_j = G_j^{(0)} + 2\mathcal{G}_j N_j^{(s)},
\end{equation}
where the factor of two originates from linearising the nonlinear saturation term $\mathcal{G}_j|\psi_j|^2\psi_j$. Neglecting the anomalous fluctuation terms, the two-mode fluctuation dynamics are governed by
\begin{equation}
	\lambda_\pm=\frac{\alpha_+-i\Delta_+}{2}
	\pm
	\sqrt{\left(\frac{\alpha_- -i\Delta_-}{2}\right)^2-J^2},
\end{equation}
with $\Delta_\pm=\Delta_1\pm\Delta_2$, and $\alpha_\pm=\alpha_1 \pm \alpha_2$. This motivates the exchange parameter
\begin{equation}
	C_\mathrm{ex}=\frac{|\alpha_-|^2+|\Delta_-|^2}{4J^2},
\end{equation}
which compares the effective non-Hermitian mismatch between the sites to coherent tunnelling. A parametric instability arises for $C_\mathrm{ex}<1$ --- such that $\lambda_\pm$ becomes complex --- while at the same time fluctuations are amplified. Indeed, Extended Data Fig.~\ref{fig:S3} shows that the lower Hopf bifurcation occurs close to $C_\mathrm{ex}=1$, showing that the adiabatically eliminated model captures the onset of oscillatory exchange very well. At larger pump powers, the upper Hopf bifurcation deviates from this simple criterion, reflecting the increasing importance of nonlinear saturation for larger photon numbers. Higher order terms in Eq.~\eqref{eq:mol_expansion} should be included to correctly capture this regime. Additionally, linearizing the nonlinear term $\mathcal G |\psi|^2\psi$ generates anomalous fluctuation terms proportional to $(\psi^{(s)})^2\delta\psi^*$, such that the full stability problem is Bogoliubov-like and involves both $\delta\psi$ and $\delta\psi^*$. The simplified exchange parameter $C_\mathrm{ex}$ and two-mode eigenvalues $\lambda_\pm$ neglect these anomalous couplings and therefore capture only the leading-order exchange instability.

\bibliography{references_all.bib}

\section*{Acknowledgments}
\noindent We thank S. Ray and R. Kramer for stimulating discussions, M. Weitz for providing laboratory infrastructure, and S.~K.~K. Rodriguez and J. Kroha for comments on the manuscript. The authors acknowledge funding from the DFG within SFB/TR 185 (277625399), Cluster of Excellence ML4Q (EXC 2004/1-390534769) and Cluster of Excellence STRUCTURES (EXC 2181/1-390900948), and from the EU (ERC, TopoGrand, 101040409). K.~P. acknowledges financial support by the DFG (Walter Benjamin programme, CONDENS, 567107833).

\section*{Author contributions}
\noindent K.~P. performed the experiments, developed the theoretical model and performed the simulations. K.~P., D.~E. and P.~S. conceived and developed the experimental apparatus. P.~S. and N.~L. imprinted and characterised the double-well mirror samples. K.~P. and J.~S. analysed the data and wrote the manuscript. All authors contributed to the discussion of the results and the manuscript.

\section*{Competing interests}
\noindent The authors declare no competing interests.

\section*{Data, code, and materials availability}
\noindent The data that support the findings of this study will be available in
the Zenodo repository.

\end{document}